\documentclass[letterpaper]{JHEP3}
\usepackage{epsfig,amsmath}

\title{Spectral Flow on the Higgs Branch and AdS/CFT Duality}

\author{
  Johanna Erdmenger, Johannes Gro\ss{}e and Zachary Guralnik \\
  Institut f\"ur Physik,
  Humboldt-Universit\"at zu Berlin,\\
  Newtonstra{\ss}e 15,
  D-12489 Berlin, Germany
  \begin{tabbing}
  Email: \= \email{jke@physik.hu-berlin.de},
            \email{jgrosse@physik.hu-berlin.de},\+\\
            \email{zack@physik.hu-berlin.de}
  \end{tabbing}
}

\preprint{HU-EP-05/09\\\hepth{0502224}}

\abstract{
We use AdS/CFT duality to study the large $N_c$ limit of the meson
spectrum on the Higgs branch of a strongly coupled, ${\cal N}=2$
supersymmetric $SU(N_c)$ gauge theory with $N_f =2$ fundamental
hypermultiplets.  In the dual supergravity description,  the Higgs
branch is described by $SU(2)$ instanton configurations on
D7-branes in an AdS background. We compute the spectral flow
parameterized by the size of a single instanton. In the large
$N_c$ limit, there is a sense in which the flow from zero to
infinite instanton size, or Higgs VEV, can be viewed as a closed
loop. We show that this flow leads to a non-trivial rearrangement
of the spectrum.
}


\newcommand{\dual}{{}^{*}}

\newcommand{\comm}[2]{\left[ #1,\, #2 \right]}

\newcommand{\spinl}{l}
\newcommand{\Nsusy}{\mathcal{N}}
\newcommand{\Afluct}{\mathcal{A}}
\newcommand{\Ainst}{A^{\text{inst}}}
\newcommand{\covD}{\mathcal{D}}
\newcommand{\Lag}{\mathcal{L}}
\newcommand{\Jcurrent}{\mathcal{J}}
\providecommand{\texorpdfstring}[2]{{#1}} 
\newcommand{\tHooft}{'t~Hooft}
\DeclareMathOperator{\tr}{tr}

\DeclareMathOperator{\im}{Im}
\DeclareMathOperator{\diag}{diag}

\begin{document}

\section{Introduction}

The spectrum of strongly coupled large $N_c$ gauge theories can in
many cases be computed using the holographically dual description
of the AdS/CFT correspondence
\cite{Maldacena:1997re,Gubser:1998bc,Witten:1998qj}. In its
original form, this duality was applicable to gauge theories with
adjoint fields only. In examples of the duality for confining
gauge theories, the spectrum of glueballs can be determined from
the spectrum of normalizable classical solutions describing small
fluctuations about a dual supergravity background
\cite{Witten:1998zw}.  Explicit computations appear in
\cite{Csaki:1998qr,Csaki:1998cb,deMelloKoch:1998qs,Brower:2000rp}.

Much effort has gone into extending AdS/CFT duality to include
theories with fundamental representations. The earliest example of
AdS/CFT duality for a theory with fundamental representations
related a conformal $\Nsusy=2$ $Sp(N)$ gauge theory to string
theory in $AdS_5 \times S^5/\mathbb{Z}_2$, with D7-branes wrapping
the $\mathbb{Z}_2$ fixed surface with geometry $AdS_5 \times S^3$
\cite{Fayyazuddin:1998fb,Aharony:1998xz}(see also
\cite{Bertolini:2001qa} for related early work).  In
\cite{Karch:2002sh}, this duality was extended to an $\Nsusy=2$
$SU(N_c)$ theory with $N_f$ massive fundamental hypermultiplets,
essentially by removing the $\mathbb{Z}_2$ orientifold, which was
justified by the fact that a probe D7-brane wrapping a
contractible $S^3$ does not lead to a tadpole requiring
cancellation.  The dual field theory is not asymptotically free,
but has a UV fixed point in the strict $N_c \rightarrow\infty$
limit.  Subsequent to this, there have also been a number of
papers generalizing the duality to confining theories with
fundamental representations
\cite{Sakai:2003wu,Nastase:2003dd,Ouyang:2003df,Nunez:2003cf,Burrington:2004id,
Erdmenger:2004dk,Hong:2004sa,Kruczenski:2004me,Paredes:2004is},
including non-supersymmetric examples in which chiral symmetry
breaking by a $\bar\Psi\Psi$ quark condensate occurs
\cite{Babington:2003vm,Babington:2003up,Kruczenski:2003uq,Barbon:2004dq,Evans:2004ia,
Ghoroku:2004sp,Bak:2004nt,DaRold:2005zs, Evans:2005ti,Sakai:2004cn}.

In this paper, we revisit the ${\cal N}=2$ theory at large $N_c$
whose AdS description was constructed in \cite{Karch:2002sh}.  The
meson spectrum of this theory at the origin of moduli space was
computed exactly in \cite{Kruczenski:2003be} by solving the
Dirac-Born-Infeld equations of motion for small fluctuations of
the D7-brane embedded in the AdS background.  Other properties of
this theory were discussed in \cite{Karch:2002xe,Hong:2003jm}. We
extend the analysis of \cite{Kruczenski:2003be} to include the
spectrum at points on the mixed Coulomb-Higgs branch. The AdS
description of these points in moduli space  has been discussed in
\cite{Guralnik:2004ve,Guralnik:2004wq,Guralnik:2005jg}, and involves
gauge field backgrounds on the embedded D7-branes with non-zero
instanton number.  We explicitly consider the part of the Higgs
branch dual to a single instanton, and compute the spectrum as a
function of the instanton size.  There is a sense in which the
zero size and infinite size limits are equivalent, modulo a
singular gauge transformation.  In the dual large $N_c$ gauge
theory,  this is an equivalence between the spectrum of the
$SU(N_c)$ theory and the $SU(N_c-1)$ theory obtained by taking the
Higgs VEV to infinity.  We shall see that the spectral flow
between these limits leads to a non-trivial re-arrangement of the
mass eigenstates and global charges. Since our purpose is to
exhibit this flow, we shall focus on a particular meson vector
multiplet with small global symmetry charges, rather than
computing the entire spectrum accessible to supergravity methods.
The specific flow we consider takes vector mesons in the $(0,0)$
representation of a global $SU(2)_L \times SU(2)_R$ symmetry,
which is unbroken at the origin of moduli space, to vector mesons
in the representation $(1,1)$.

The organization of this paper is as follows.  In section 2, we
review the AdS description of the Higgs branch of the $\Nsusy=2$
theory with fundamental hypermultiplets described in
\cite{Karch:2002sh}.  Section 3 contains a review of the AdS
description of the Higgs branch and parts of the mixed
Coulomb-Higgs branch.  In section 4, we discuss the AdS/CFT
dictionary at the points on the moduli space dual to a single
instanton.  In section 5, we compute the meson spectrum by solving
for classical fluctuations about the instanton background.

\section{SUGRA dual of an \texorpdfstring{$\Nsusy=2$}{N=2} theory with fundamental representations}

The specific $\Nsusy=2$ gauge theory which we consider is dual to
string theory in $AdS_5 \times S^5$ with $N_f$ D7-branes wrapping
a surface which is asymptotically $AdS_5 \times S^3$. This
particular duality was originally described in
\cite{Karch:2002sh}. The matter content of this gauge theory is
that of the $\Nsusy=4$ $SU(N_c)$ gauge theory, with an added $N_f$
(possibly massive) fundamental hypermultiplets. In ${\cal N}=1$
superspace, the Lagrangian is
\begin{equation}
\begin{split}
\Lag 
  = \im & \left[\tau \int d^2 \theta d^2 \bar\theta
         \left(\tr (\bar \Phi_I e^V \Phi_I e^{-V}) 
          + Q_i^\dagger e^V Q^i + \tilde Q_i e^{V} \tilde Q^{i\dagger} \right) 
          \right. \\
  & \left. + \, \tau \int d^2 \theta \left(\tr ({\cal W}^\alpha{\cal W}_\alpha) + W\right) \, +
            \tau \int d^2 \bar \theta \left(\tr (\bar{\cal W}_{\dot \alpha}
\bar{\cal W}^{\dot\alpha}) + \bar W\right) \,  \right]
\end{split}
\end{equation}
where the superpotential $W$ is
\begin{align}
W= \tr (\epsilon_{IJK}\Phi_I\Phi_J\Phi_K) + \tilde Q_i (m +
\Phi_3) Q^i
\end{align}
The superfields $Q^i$ and $\tilde Q_i$, labeled by the flavor
index  $i=1 \cdots N_f$, make up the ${\cal N}=2$ fundamental
hypermultiplets.

At finite $N_c$ this theory is not asymptotically free,  and the
corresponding string background suffers from an uncancelled
tadpole.  However, as in \cite{Karch:2002sh}, we focus strictly on
the $N_c\rightarrow\infty$ limit with fixed $N_f$. In this case
the beta function for the \tHooft{} coupling vanishes, and the
dual AdS string background has no tadpole problem because the
probe D7-branes wrap a contractible $S^3$. Although contractible,
the background is stable.  The tachyonic mode associated with
shrinking the $S^3$ satisfies (saturates) the
Brei\-ten\-loh\-ner-Freed\-man bound \cite{Breitenlohner:1982jf}. Furthermore
the $AdS_5 \times S^3$ embedding has been shown to be
supersymmetric \cite{Skenderis:2002vf}.

\TABLE{%
  \parbox{0.9\linewidth}{\center 
  \epsfig{file=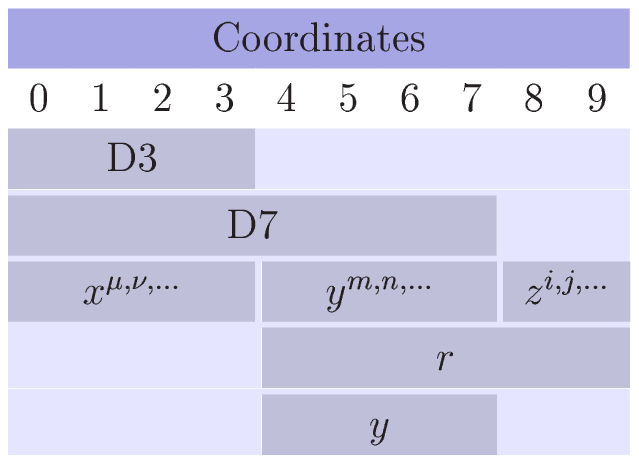}
  \caption{\label{tab:indices}Index conventions}
}}

The $AdS_5 \times S^5$ background is given by
\begin{align}\label{AdS}
    ds^2 &= \makebox[0pt][l]{$H^{-1/2}(r) \eta_{\mu\nu}dx^\mu dx^\nu +%
           H^{1/2}(r) ( d\vec{y}^{\,2} + d\vec{z}^{\,2} ),$} \nonumber\\
    H(r) &= \frac{L^4}{r^4},       & r^2 &= \vec{y}^{\,2} + \vec{z}^{\,2}, \nonumber\\
    L^4 &= 4\pi g_s N_c (\alpha')^2, & \vec{y}^{\,2} &= \sum_{m=4}^{7} y^m y^m, \\
    C^{(4)}_{0123} &= H^{-1} ,     & \vec{z}^{\,2} &= (z^8)^2+(z^9)^2, \nonumber\\[2ex]
    e^{\phi}&=e^{\phi_\infty}=g_s. \nonumber
\end{align}

We have only written the components of the Ramond-Ramond four-form
which will be relevant to our computations. $N_f$ D7-branes are
embedded in this geometry according to
\begin{align}\label{embed}
  z^8 &= 0, &
  z^9 &= (2\pi\alpha')m \, .
\end{align}
This
leads to the induced metric
\begin{align}\label{D7geom}
\begin{split}
  ds^2_{\text{D7}} &=
     H^{-1/2}(r) \eta_{\mu\nu}dx^\mu dx^\nu +
     H^{1/2}(r) d\vec{y}^{\,2}, \\[1ex]
  r^2 &= y^2 +  (2\pi\alpha')^2 m^2, \qquad y^2 \equiv y^m y^m.
\end{split}
\end{align}
The parameter $m$ corresponds to the mass of the fundamental
hypermultiplets in the dual $\Nsusy=2$ theory.  For $m=0$, the
geometry (\ref{D7geom}) is $AdS_5 \times S^3$, while for $m\neq
0$, the geometry approaches $AdS^5 \times S^3$ at large $r$. As
long as $N_f$ is held fixed in limit $N_c\rightarrow\infty$ with
fixed $\lambda = g_s N_c \gg 1$, there is no need to consider the
back-reaction of the D7-branes on the bulk geometry.

\section{The Higgs branch}
\subsection{Field Theory \label{sect:higgs-ft}}
The $\Nsusy=2$ theory  to which the supergravity background
\eqref{AdS}--\eqref{D7geom} is dual contains an ${\cal N}=2$
vector multiplet, one adjoint hypermultiplet, and $N_f$
fundamental hypermultiplets. In ${\cal N}=1$ language, the
superpotential is
\begin{align}
  W= \tilde Q_i(m + \Phi_3) Q^i + \tr [\Phi_1,\Phi_2]\Phi_3\, .
\end{align} The chiral superfields $Q^i$ and $ \tilde Q_i$ belong
to the fundamental hypermultiplets, with flavor index $i$.
$\Phi_1$ and $\Phi_2$ belong to an adjoint hypermultiplet, while
$\Phi_3$ belongs to the vector multiplet. We denote the (scalar)
bottom components of the superfields by lowercase letters. On the
Higgs branch, the vector multiplet moduli $\phi_3$ vanish while
$q^i$ and $\tilde q_i$ have non-zero expectation
values\footnote{The scalars $\phi_1$ and $\phi_2$ belonging to the
adjoint hypermultiplet may also have non-zero expectation values
on the Higgs branch.}. There are also mixed Coulomb-Higgs vacua,
for which both $q^i,\tilde q_i$ and $\phi_3$ have non-zero
expectation values.

For non-zero $m$ and vanishing $\phi_3$,  the fundamental
hypermultiplets are massive and there is no Higgs branch.  However
there is a mixed Coulomb-Higgs branch when $\phi_3$ has an
expectation value such that some components of the hypermultiplets
are massless. An example of a point on a mixed Coulomb-Higgs
branch is given by a diagonal $\phi_3$ for which all but the last
$k$ entries are vanishing:
\begin{align}\label{Coul}
  \phi_3 = \begin{pmatrix} 0& & & & & \cr &\ddots & & & &  \cr
        & &0 & & & \cr & & &-m & &  \cr & & & &\ddots & \cr
        & & & & & -m
\end{pmatrix}\, .
\end{align}
In this case, the F-flatness equations $\tilde q_i (\phi_3+m) =
(\phi_3+m) q^i = 0$ permit fundamental hypermultiplet expectation
values in which only the last $k$ entries of $q^i$ and $\tilde
q_i$ are non-zero;
\begin{align}\label{cond}q^i= \begin{pmatrix} 0 \cr \vdots \cr 0 \cr \alpha_1^i \cr
\vdots \cr \alpha_k^i \end{pmatrix}\, , \qquad  \tilde q_i =
\begin{pmatrix} 0 & \cdots & 0 & \beta_{1i} & \cdots
& \beta_{ki} \end{pmatrix}\, .\end{align}  There are additional F
and D-flatness constraints  which we have not explicitly written.

In string theory, nonzero
entries in (\ref{cond}) physically correspond to D3-branes
which are coincident with and dissolved within the D7-branes.
Dissolved D3-branes can be viewed as instantons in the
eight-dimensional world-volume theory on the D7-branes
\cite{Douglas:1995bn}, due to the Wess-Zumino coupling
\begin{equation}
S_{WZ} = \frac{T_7(2\pi\alpha')^2}{4} \int C|_{PB}^{(4)}
   \wedge \tr (F\wedge F) \, .
\end{equation}
In fact, there is a known exact map between the moduli space of
Yang-Mills instantons and the Higgs branch of the $p+1$
dimensional theory arising on the Dp -- Dp+4 intersections.  The
ADHM constraints from which instantons are constructed are
equivalent to the F and D-flatness equations
of the $p+1$ dimensional theory \cite{Witten:1995gx,Douglas:1996uz}
 (see also \cite{Dorey:2002ik} for a review).  The existence of
instanton solutions for the D7-brane embedded in (\ref{AdS})
according to (\ref{embed}) is a non-trivial consequence of AdS/CFT
duality \cite{Guralnik:2004ve,Guralnik:2004wq,Guralnik:2005jg}.

\subsection{Supergravity description of the Higgs branch}

The AdS/CFT dictionary relates the fundamental hypermultiplets of
the ${\cal N}=2$ theory to degrees of freedom on D7-branes
embedded in the AdS geometry according to (\ref{embed}).  In light
of the one-one correspondence between instantons and the Higgs
branch,  one expects that the supergravity description of the
Higgs branch involves instanton solutions of the non-Abelian
Dirac-Born-Infeld action which describes the  D7 branes.

The effective action describing D7-branes in a curved background
is
\begin{align}\label{ac}
S &= T_7 \int  \sum_r C^{(r)} \wedge \tr e^{2\pi  \alpha'
F}\nonumber\\
&+ T_7\int\,d^{8}\xi\, \sqrt{g} e^{-\phi}(2\pi
\alpha')^2\frac{1}{2} \tr \left( F_{\alpha\beta}F^{\alpha\beta}
\right) + \cdots \, , 
\end{align} where we have not written terms involving fermions and scalars.
This action is the sum of a Wess-Zumino term, a Yang-Mills term,
and an infinite number of corrections at higher orders in
$\alpha'$ indicated by $\cdots$ in \eqref{ac}. Since we need to
consider at least two flavors (two D7's) in order to have a Higgs
branch, the DBI action is non-Abelian. The correspondence between
instantons and the Higgs branch suggests that the equations of
motion should be solved by field strengths which are self-dual
with respect to a flat four-dimensional metric.  In this paper, we
work to leading order only in the large \tHooft{} coupling
expansion generated by AdS/CFT duality, which allows one to only
consider the leading term in the $\alpha'$ expansion of the
action.  Constraints on unknown higher order terms arising from
the existence of instanton solutions, as well as the exactly known
metric on the Higgs branch, were discussed in
\cite{Guralnik:2004ve,Guralnik:2005jg}.

At leading order in $\alpha'$, field strengths which are self dual
with respect to the flat four dimensional metric $ds^2 =
\sum_{m=1}^4 dy^mdy^m$ solve the equations of motion, due to a
conspiracy between the Wess-Zumino and Yang-Mills term. Inserting
the explicit AdS background values \eqref{AdS} for the metric and
Ramond-Ramond four-form into the action for D7-branes embedded
according to \eqref{embed},  with non-trivial field strengths only
in the directions $y^m$,  gives
\begin{align}\label{theactn}
\begin{split}
  S &= \frac{T_7 (2\pi\alpha')^2}{4} \int\, d^4x\,d^4y\, H(r)^{-1} \,
       \left(-\frac{1}{2}\epsilon_{mnrs}F_{mn}F_{rs} +
       F_{mn}F_{mn}\right) = \\
    &= \frac{T_7 (2\pi\alpha')^2}{2} \int \, d^4x\,d^4y\,  H(r)^{-1} F_-^2 \, ,
\end{split}
\end{align}
where $r^2 = y^m y^m + (2\pi\alpha' m)^2$ and  $F^-_{mn} =
\frac{1}{2}(F_{mn}-\frac{1}{2}\epsilon_{mnrs}F_{rs})$. Field
strengths $F^-_{mn} = 0$, which are self-dual with respect to the
flat metric $dy^mdy^m$, manifestly solve the equations of motion.
These solutions correspond to points on the Higgs branch of the
dual $\Nsusy=2$ theory.  Strictly speaking, this is a point on the
mixed Coulomb-Higgs branch if $m \ne 0$,  with expectation values
of the form \eqref{Coul}, \eqref{cond}. We emphasize that in order
to neglect the back-reaction due to dissolved D3-branes, we are
considering a portion of the moduli space for which the instanton
number $k$ is fixed in the large $N_c$ limit.

\section{Higgs branch AdS/CFT dictionary}

For simplicity, we  consider the case $N_f=2$, corresponding to
two D7-branes,  which is the minimum value giving a non-trivial
Higgs branch. For $m = 0$, the AdS geometry \eqref{AdS} together
with the embedding \eqref{embed}, \eqref{D7geom},  is invariant
under $SO(2,4) \times SU(2)_L \times SU(2)_R \times U(1)_R \times
SU(2)_f$. The combination $SU(2)_L \times SU(2)_R$ acts as $SO(4)$
rotations of the coordinates $y^m$.  The $SO(2,4)$ factor is the
conformal symmetry of the dual gauge theory. The $SU(2)_L$ factor
corresponds to a global symmetry of the dual gauge theory, while
$SU(2)_R \times U(1)_R$ corresponds to the R symmetries. Finally
$SU(2)_f$ is the gauge symmetry of the two coincident D7-branes
which, at the AdS boundary, corresponds to the flavor symmetry of
the dual gauge theory.

For $m\ne 0$,  the symmetry is broken to $SO(1,3) \times SU(2)_L
\times SU(2)_R \times SU(2)_f$.  This is broken further if there
is an instanton background on the D7-branes. We  focus on that
part of the Higgs branch, or mixed Coulomb-Higgs branch, which is
dual to a single instanton centered at the origin $y^m=0$. The
instanton, in ``singular gauge,'' is given by
\begin{align}\label{thinst}
A_\mu = 0,\qquad A_m = \frac{2\Lambda^2
\bar\sigma_{nm}y_n}{y^2(y^2 + \Lambda^2)}
\end{align}
where $\Lambda$ is the instanton size, and
\begin{align} 
  \bar\sigma_{mn} &\equiv \frac{1}{4}(\bar\sigma_m \sigma_n - \bar\sigma_n \sigma_m) \, , &
  \sigma_m &\equiv (i\vec\tau, 1_{2\times 2})\, , \nonumber \\
  \sigma_{mn}  & \equiv \frac{1}{4}(\sigma_m \bar\sigma_n - \sigma_n \bar\sigma_m)\, , &
  \bar\sigma_m & \equiv \sigma_m^{\dagger} = (-i\vec\tau, 1_{2\times2})\,.
\end{align}
with $\vec\tau$ being the three Pauli-matrices. We choose singular
gauge, as opposed to the regular gauge in which $A_n = 2
\sigma_{mn}y^m/(y^2 + \Lambda^2)$,  because of the improved
asymptotic behavior at large $y$. In the AdS setting,  the Higgs
branch should correspond to a normalizable deformation of the
background at the origin of the moduli space.  The singularity of
\eqref{thinst} at $y^m =0$ is not problematic for computations of
physical (gauge invariant) quantities.

 The instanton \eqref{thinst} breaks the symmetries to
\begin{align}G = SO(1,3) \times SU(2)_L \times \diag (SU(2)_R
\times SU(2)_f)\, ,\end{align} and corresponds to a point on the
Higgs branch
\begin{gather}
q_{i \alpha }  =  \, v \, \varepsilon_{i \alpha
} \, ,  \qquad v = \frac{\Lambda}{2\pi \alpha'} \, ,
\end{gather}
where $q_{i \alpha  }$ are scalar components of the fundamental
hypermultiplets, labeled by a $SU(2)_f$ index $i=1,2$, and a
$SU(2)_R$ index $\alpha =1,2$.  All the broken symmetries are
restored in the ultraviolet (large $r$), where the theory becomes
conformal.

\section{Fluctuation spectrum}

To determine the spectrum on the Higgs branch, we now consider
fluctuations about the instanton background \eqref{thinst}.  There
are some obvious fluctuations of $A_m$ dual to massless scalar
mesons, namely fluctuations corresponding to changes in the
instanton moduli.  We  focus instead on fluctuations of
$A_\mu$ in the lowest representations of the un-broken $SU(2)_L
\times \diag (SU(2)_R \times SU(2)_f)$ which are dual to
vector mesons, as well as the scalar fluctuations belonging to the
same super-multiplet.

In terms of coordinates $x^\mu, y^m$ on the D7-brane world-volume,
with the former corresponding to the space-time directions of the
dual gauge theory,  the D7-brane action is
\begin{align} \label{eqn:gaugequadraticaction}
   S_{D7} = & \tfrac{(2\pi\alpha')^2 T_7}{4}
   \int d^4x\, d^4y\; \tr \biggl[
             H(r) F_{\mu\nu} F_{\mu\nu}
              + 2 F_{m\nu}F_{m\nu} \biggr.   \nonumber\\
            & \hspace{4cm} 
\biggl.   \, + \, H^{-1}(r) \, \tfrac{1}{2} (F_{mn}- \dual F_{mn})
              ( F_{mn} - \dual F_{mn} ) \biggr] \, , \hfill
\end{align}
with $r^2 = y^m y^m + (2\pi\alpha'm)^2$. We have excluded scalar
and fermionic terms, as well as higher dimension operators which
give rise to subleading corrections in the large 't Hooft coupling
expansion. The equations of motion for this action are
\begin{subequations}
\begin{align}
 H(r)\covD_\mu F_{\mu\nu}  + \covD_m F_{m\nu} &= 0,  \label{eqn:eomgreekexpanded}\\
  \covD_\mu F_{\mu n} + \covD_m \left[ H^{-1}(r) (F_{mn} - \dual F_{mn}) \right] &=0.
\end{align}
\end{subequations}
Let us consider the equations of motion for small fluctuation
$\Afluct$ about the instanton, defined by $\Afluct \equiv A - \Ainst$.
The normalizable solutions of the classical equations for
fluctuations $\Afluct_\mu$ with $\Afluct_m=0$ determine the
spectrum of vector mesons. To linear order in $\Afluct_\mu$, the
equations of motion are
\begin{subequations}
\begin{align}
  \covD_n (\partial_\mu \Afluct_\mu) &= 0, \label{eqn:transv}\\
  H \partial_\mu \partial_\mu \Afluct_\nu
  + \partial_m \partial_m \Afluct_\nu +
\partial_m \comm{ \Ainst_m }{ \Afluct_{\nu} } \hspace{2.1cm} &\nonumber \\
  +  \comm{ \Ainst_m }{ \partial_m \Afluct_\nu }
  +  \comm{ \Ainst_m }{ \comm{ \Ainst_m }{ \Afluct_{\nu} } }
 &= 0. \label{eqn:flucteom}
\end{align}
\end{subequations}
The simplest (non-Abelian) ansatz for the fluctuations ${\cal
A}_\mu$ is given by
\begin{gather} \label{ansatz}
  \Afluct_\mu{}^{(a)} = \xi_\mu(k) f(y) e^{ik_\mu x_\mu} \tau^a \, ,
  \qquad y^2 \equiv y^m y^m\,  ,
\end{gather}
which is a singlet under $SU(2)_L$ and a triplet under 
$\diag(SU(2)_R\times SU(2)_f)$. Equation (\ref{eqn:transv}) is
solved by  $k_\mu \xi_\mu =0$, while (\ref{eqn:flucteom}) becomes
\begin{align}
  0 = \biggl[ \frac{M^2 L^4}{(y^2 + (2\pi\alpha')^2m^2)^2}
                - \frac{8 \Lambda^4}{y^2(y^2+\Lambda^2)^2}
                + \frac{1}{y^3}\partial_y (y^3 \partial_y ) \biggr] f(y).
                \label{eqn:ansatzeom}
\end{align}
where $M^2 = -k_\mu k_\mu$.  To determine the spectrum,  we must
find the values of $M^2$ for which this equation admits
normalizable solutions.

It is convenient to work in units of the hypermultiplet mass by
defining
\begin{align}
        \tilde y &\equiv \frac{y}{2\pi\alpha'm}, &
        \tilde\Lambda &\equiv \frac{\Lambda}{2\pi\alpha'm}, &
        {\tilde M}^2 &\equiv \frac{M^2 L^4}{(2\pi\alpha' m)^2},
\end{align}
such that equation \eqref{eqn:ansatzeom} becomes
\begin{align}
  0 = \biggl[ \frac{{\tilde M}^2}{({\tilde y}^2 + 1)^2}
                - \frac{8 {\tilde\Lambda}^4}{{\tilde y}^2({\tilde y}^2
                +{\tilde\Lambda}^2)^2}
                + \frac{1}{{\tilde y}^3}\partial_{\tilde y}
                ({\tilde y}^3 \partial_{\tilde y} ) \biggr] f({\tilde y}).
                \label{eqn:eomrescaled}
\end{align}
This equation can be solved analytically in the limits of zero or
infinite instanton size. For finite $\tilde\Lambda$ we solve it
numerically.

At large  $\tilde y$, (\ref{eqn:eomrescaled}) has two linear
independent solutions whose asymptotics are given by
$\tilde y^{-\lambda}$ with $\lambda=0,2$.  The normalizable
solutions corresponding to vector meson states behave as
$\tilde y^{-2}$ asymptotically. From the standard AdS/CFT
correspondence, one expects $\lambda=\Delta$ and
$\lambda=4-\Delta$, where $\Delta$ is the UV conformal dimension
of the lowest dimension operator with the quantum numbers of the
vector meson.  However, the kinetic term does not have a standard
normalization, i.e.~the radial component of the Laplace operator
appearing in the equation above is not (only)
$\partial_{\tilde y}^2$, and consequently an extra factor of
$\tilde y^\alpha$, for some $\alpha$, appears in the expected
behaviour; so we have $\lambda=\alpha+\Delta$, $\alpha+4-\Delta$.
From the difference we conclude that $\Delta=3$. The dimensions
and quantum numbers are those of the $SU(2)$ flavor current,
\begin{gather} \label{flavour}
  {\cal J}_\mu^b
    = - \bar \psi^{\pm i} \gamma_\mu \sigma^b{}_{ij} \psi_{\mp}{}^{j}
    + \bar q^{\alpha i } \stackrel{\leftrightarrow}{D}_\mu
           \sigma^b{}_{ij}\, q_\alpha{}^j \, ,
\end{gather}
with $\alpha$ the $SU(2)_R$ index and $i,j$ the flavor indices.
This current has $SU(2)_R\times SU(2)_L \times U(1)$  quantum
numbers $(0,0)_0$.

The asymptotic behavior of the supergravity solution is
\begin{align}
A^{\mu}_{b (a) } =
\xi^{\mu}(k)e^{ik_{\mu}x^{\mu}}f(\tilde y)\delta_{ab} \sim
\tilde y^{-2} \langle a,\xi,k| \Jcurrent^{\mu}_b(x) |0 \rangle \, ,
\end{align}
where $\Jcurrent^{\mu}$ is the SU(2) flavor current and $|a,\xi,k\rangle$ is a
vector meson with polarization $\xi$, momentum $k$, and flavor
triplet label $a$. Note that the index $b$ in $A^{\mu}_{b (a) }$
is a Lie algebra index, whereas the index $(a)$ labels the flavor
triplet of solutions.

For small radii $\tilde y$, the asymptotics of the general
solution of \eqref{eqn:eomrescaled} is $c_1 \tilde y^2 + c_2
\tilde y^{-4}$.   Requiring a normalizable solution with regular
behavior at the origin gives a discrete spectrum of allowed
$\tilde M^2$. We have calculated $\tilde M^2$ for the lowest-lying
modes using a standard numerical shooting method. The results are
shown in figure \ref{fig:mesonmass}.

Because of ${\cal N}=2$ supersymmetry, there will be a number of
other fluctuations with the same spectrum as the vector mesons,
which are required to fill out massive ${\cal N}=2$ vector
multiplets. These include, for example, fluctuations of the
adjoint scalars on the D7-brane. At quadratic order, the part of
the D7-brane action involving these scalars and the corresponding
equation of motion is
\begin{gather}
    S_{D7} = - \tfrac{(2\pi\alpha')^2 T_7}{4}
    \int d^4x\, d^4y\; \tr \biggl[
               H(r) \covD_\mu \Phi \covD_\mu \bar\Phi
               + \covD_m \Phi \covD_m \bar \Phi \biggr]
          \label{eqn:scalarquadraticaction}, \\
   H(r) \partial_\mu \partial_\mu \Phi + \covD_m \covD_m \Phi = 0, \label{eqn:scalareom}
\end{gather}
where
\begin{align}
  \covD_m \covD_m \Phi &= \partial_m\partial_m \phi +  \comm{\Ainst_m}{\partial_m \Phi} +
                     \partial_m \comm{\Ainst_m}{\Phi} \\
                    &\quad +  \comm{\Ainst_m}{\comm{\Ainst_m}{\Phi} }.
\end{align}
As \eqref{eqn:scalareom} coincides with the equation of motion for
the gauge field \eqref{eqn:flucteom}, except for the vector index
present, the ansatz for the scalar field
\begin{equation} \label{scalaransatz}
  \Phi = f(\tilde y)e^{i k_\mu x_\mu}  \tau^a
\end{equation}
yields exactly the same differential equation
\eqref{eqn:ansatzeom}.

The scalar fluctuations (\ref{scalaransatz}) are dual to  the
descendant $QQ(q_i \bar q^i)$ of the scalar bilinear $q_i \bar
q^i$, which  has conformal dimension $\Delta=3$. At $\Lambda =0$
the scalar bilinear is in the $(0,0)$ representation of the
unbroken $SU(2)_L\times SU(2)_R$ symmetry.

\subsection{Spectral flow}

In the limits of zero or infinite instanton size, the equations
\eqref{eqn:eomrescaled} become \begin{align}
  0 = \biggl[ \frac{{\tilde M}^2}{({\tilde y}^2 + 1)^2}
                - \frac{\spinl (\spinl +2 )}{{\tilde y}^2}
                + \frac{1}{{\tilde y}^3}\partial_{\tilde y}
                ({\tilde y}^3 \partial_{\tilde y} ) \biggr]
  f({\tilde y}), \label{asy}
\end{align} with $l = 0,2$ for zero or infinite $\tilde \Lambda$
respectively.  Note that these are the same equations found in
\cite{Kruczenski:2003be} for fluctuations about the trivial
background $A_\mu = A_m = 0$ of the form
\begin{align}
{\cal A}^{\mu} = \xi^{\mu}(k)e^{ik_{\mu}x^{\mu}}f(y) {\cal
Y}_l(S^3) \, ,
\end{align}
where ${\cal Y}_l(S^3)$ are spherical harmonics on $S^3$, which
transform as $(\frac{l}{2},\frac{l}{2})$ representations under
$SU(2)_L \times SU(2)_R$.  Regular, normalizable solutions exist
for \cite{Kruczenski:2003be}
\begin{align} \label{spectrumM}
  {\tilde M}^2 & = 4(n+\spinl+1)(n+\spinl+2)\, , \qquad
\end{align}
for integer $n\ge 0$. The spectra computed in
\cite{Kruczenski:2003be} are valid for a single flavor, for which
there is no Higgs branch, or for multiple flavors at the origin of
moduli space.

In the limit of infinite instanton size, the field strength
vanishes locally, and one would again expect that the spectrum is
the same as that at the origin of moduli space.  In the dual gauge
theory, the infinite size limit corresponds to the limit of
infinite Higgs VEV, which reduces the gauge group from $SU(N_c)$
to $SU(N_c-1)$. The distinction between $SU(N_c)$ and $SU(N_c-1)$
is negligible in the large $N_c$ limit. Thus, in some sense,  a
flow from zero to infinite size is a closed loop beginning and
ending at the origin of moduli space.  The spectrum is the same,
although it could rearrange itself in a non-trivial way during the
flow. This is indeed the case, as can be seen from figure
\ref{fig:mesonmass}(b).  We shall argue below that this flow takes
vector mesons in the $(0,0)$ representation of the global $SU(2)_L
\times SU(2)_R$ symmetry, which is unbroken at the origin of the
moduli space, to vector mesons in the representation $(1,1)$.

\FIGURE{
  \includegraphics[width=0.7\linewidth]{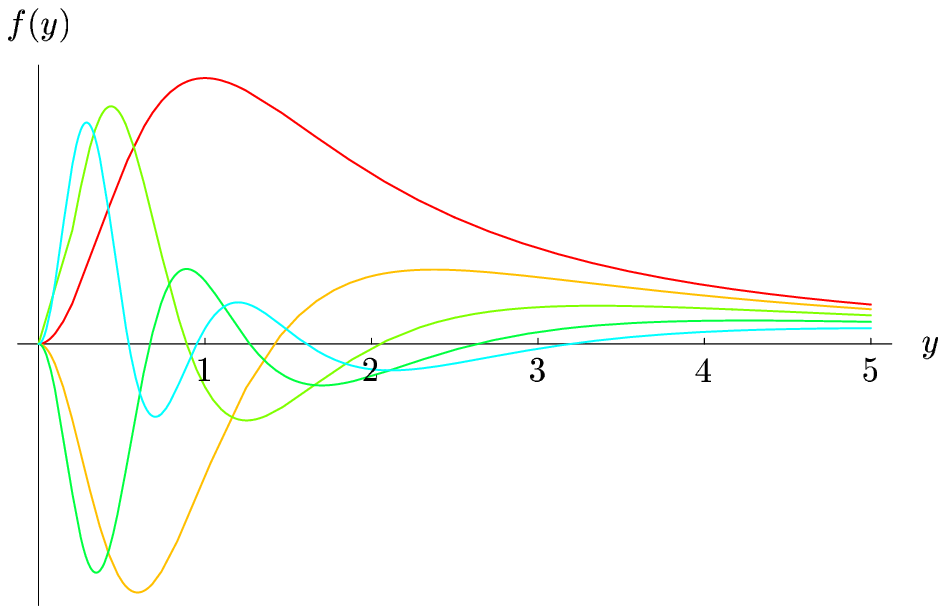}\\[1ex]
  {\small
  \begin{tabular}{r@{ }p{0.6\linewidth}}
    (a)& Regular solutions of \protect\eqref{eqn:eomrescaled}. The $f$ axis
         has arbitrary scale.
  \end{tabular}
  }\\[2ex]
  \includegraphics[width=0.75\linewidth]{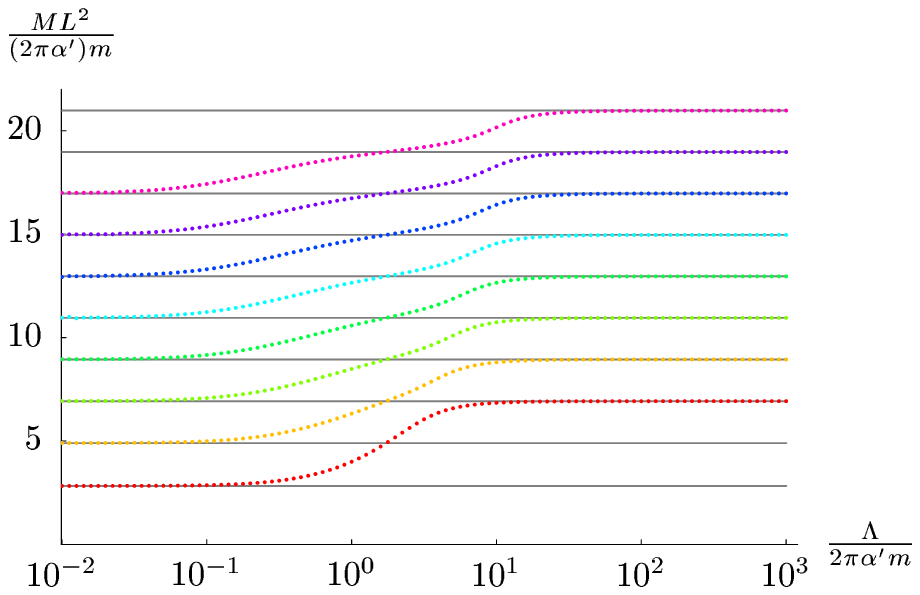}\\[1ex]
  {\small
  \begin{tabular}{r@{ }p{0.6\linewidth}}
    (b)& Numerically determined meson masses as function of the
         Higgs VEV. 
  \end{tabular}
  }\\[1ex]
\caption[Meson Masses]{\label{fig:mesonmass}
 Each dotted line represents a regular solution of the equation of
 motion, corresponding to a vector multiplet of mesons.  The
 vertical axis in (b)
 is $\sqrt{\lambda}M/m$ where $M$ is the meson mass, $\lambda$
 the 't~Hooft coupling and $m$ the quark mass. The horizontal axis
 is $v/m$ where $v= \Lambda/2\pi \alpha' $ is the Higgs VEV.
 In the limits of zero and infinite instanton size (Higgs VEV), one
        recovers the spectrum (gray horizontal lines)
        obtained analytically in the absence of an instanton background by
        \protect\cite{Kruczenski:2003be}.
        }
}

In singular gauge,  the infinite size instanton is given by
\begin{align}
A_n = 2 \frac{\bar\sigma_{mn}y^m}{y^2}
\end{align}
such that the ansatz \eqref{ansatz} gives \eqref{asy} with $l=2$.
Note however that the solution of the form \eqref{ansatz} involves
the trivial spherical harmonic (a constant) on $S^3$. Naively, it
would seem impossible for a flow between zero and infinite
instanton size to generate an $l=2$ spherical harmonic on $S^3$,
starting with the trivial $l=0$ harmonic,  since $SU(2)_L$ is
unbroken by the instanton \eqref{thinst}. However, in the limit of
infinite size, the correspondence with the the spectrum computed
in \cite{Kruczenski:2003be} at the origin of moduli space is
apparent only after making the gauge transformation
\begin{align}\label{gtrans}U = \sigma^m y^m/|y|\, ,\end{align}
which gives\footnote{The instanton number is non-zero, so we can
only get vanishing $F_{mn} =0$ by taking the instanton size to
infinity at fixed $y^m$.  The AdS wavefunctions associated with
the spectrum we are discussing are localized in a region which is
fixed as the instanton size goes to infinity.} $A_n=0$. Acting on
the ansatz \eqref{ansatz}, this gauge transformation gives
\begin{align}
{\cal A}_\mu{}^{(a)} = \xi_\mu(k) f(y)
 e^{ik_\mu x_\mu} \hat y^m \hat y^n \sigma^m\tau^a\bar\sigma^n  \, ,
\end{align}
with $\hat y^m = y^m/|y|$. The matrix elements of ${\cal A_\mu}$
contain only $l=2$ spherical harmonics on $S^3$,  since $\hat
y^n\hat y^m$ multiplies $C^{nm} \equiv \sigma^m \tau^a \bar
\sigma^n$ which has zero trace: $C^{mm} = \sigma^m \tau^a \bar
\sigma^m = 0$. Note that the singular gauge transformation
\eqref{gtrans} is large\footnote{This gauge transformation relates
the ``singular gauge'' to the ``regular gauge'' (see for instance
\cite{Dorey:2002ik}).} and does not leave physical states or the
$SU(2)_L \times SU(2)_R \times SU(2)_f$ global charges invariant.

\EPSFIGURE{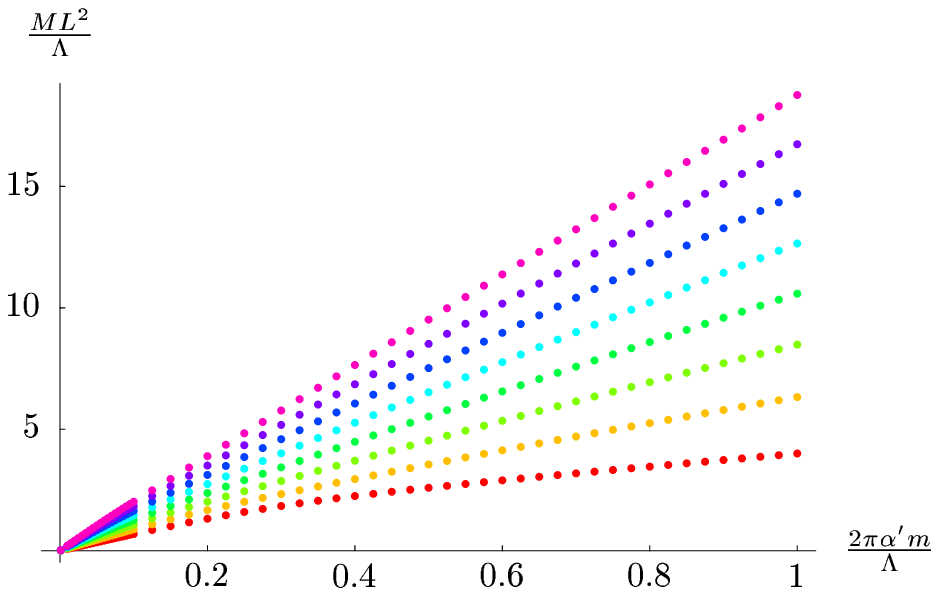,width=0.75\linewidth}{\label{fig:mesonmass2}
        Numerical results for the meson mass spectrum as function of the quark
        mass. Both for $m/\Lambda \rightarrow 0$ and for 
        $m/\Lambda\rightarrow \infty$,
        the curves become linear, however with different slope. The
        asymptotic slopes correspond to the constant values approached 
        in figure
        \protect\ref{fig:mesonmass}(b).}

\section{Conclusions}

We have used AdS/CFT duality to compute meson spectra on the Higgs
branch of a strongly coupled  ${\cal N}=2$ gauge theory at large
$N_c$. The Higgs branch is dual to instanton configurations on
D7-branes embedded in an AdS geometry. The meson spectrum is
determined by finding the normalizable solutions describing small
fluctuations about this background. We have focused on a
particular portion of the mixed-Coulomb Higgs branch which is dual
to a single instanton centered at the origin. Our intention has
not been to compute the full spectrum accessible to supergravity
calculations, but rather to illustrate a non-trivial spectral flow
as one varies the size of the instanton between zero and infinity.
To this end we have focused on a particular ansatz for solutions
dual to vector mesons with the smallest global symmetry charges.
In the large $N_c$ limit, there is a sense in which the flow from
zero to infinite instanton size, or Higgs VEV, is a
non-contractible closed loop. Although the spectrum at the
endpoints is the same, the flow leads to a non-trivial
rearrangement of global symmetry charges and mass eigenstates.

\acknowledgments
We thank R. Easther, J. Maldacena and D. Mateos for useful
discussions. The research of J.E. and Z.G. is supported by the DFG
(Deutsche Forschungsgemeinschaft) within  the Emmy Noether
programme, grant ER301/1-4. \newline J.G. acknowledges support through the
DFG Graduiertenkolleg ``The Standard Model of Particle Physics --
structure, precision tests and extensions''.

\bibliographystyle{JHEP}
\bibliography{specflow-jhep}

\providecommand{\href}[2]{#2}\begingroup\raggedright\begin{thebibliography}{10}

\bibitem{Maldacena:1997re}
J.~M. Maldacena, {\it The large n limit of superconformal field theories and
  supergravity},  {\em Adv. Theor. Math. Phys.} {\bf 2} (1998) 231--252,
  [\href{http://xxx.lanl.gov/abs/hep-th/9711200}{{\tt hep-th/9711200}}].

\bibitem{Gubser:1998bc}
S.~S. Gubser, I.~R. Klebanov, and A.~M. Polyakov, {\it Gauge theory correlators
  from non-critical string theory},  {\em Phys. Lett.} {\bf B428} (1998)
  105--114, [\href{http://xxx.lanl.gov/abs/hep-th/9802109}{{\tt
  hep-th/9802109}}].

\bibitem{Witten:1998qj}
E.~Witten, {\it Anti-de sitter space and holography},  {\em Adv. Theor. Math.
  Phys.} {\bf 2} (1998) 253--291,
  [\href{http://xxx.lanl.gov/abs/hep-th/9802150}{{\tt hep-th/9802150}}].

\bibitem{Witten:1998zw}
E.~Witten, {\it Anti-de sitter space, thermal phase transition, and confinement
  in gauge theories},  {\em Adv. Theor. Math. Phys.} {\bf 2} (1998) 505--532,
  [\href{http://xxx.lanl.gov/abs/hep-th/9803131}{{\tt hep-th/9803131}}].

\bibitem{Csaki:1998qr}
C.~Csaki, H.~Ooguri, Y.~Oz, and J.~Terning, {\it Glueball mass spectrum from
  supergravity},  {\em JHEP} {\bf 01} (1999) 017,
  [\href{http://xxx.lanl.gov/abs/hep-th/9806021}{{\tt hep-th/9806021}}].

\bibitem{Csaki:1998cb}
C.~Csaki, Y.~Oz, J.~Russo, and J.~Terning, {\it Large n {QCD} from rotating
  branes},  {\em Phys. Rev.} {\bf D59} (1999) 065012,
  [\href{http://xxx.lanl.gov/abs/hep-th/9810186}{{\tt hep-th/9810186}}].

\bibitem{deMelloKoch:1998qs}
R.~de~Mello~Koch, A.~Jevicki, M.~Mihailescu, and J.~P. Nunes, {\it Evaluation
  of glueball masses from supergravity},  {\em Phys. Rev.} {\bf D58} (1998)
  105009, [\href{http://xxx.lanl.gov/abs/hep-th/9806125}{{\tt
  hep-th/9806125}}].

\bibitem{Brower:2000rp}
R.~C. Brower, S.~D. Mathur, and C.-I. Tan, {\it Glueball spectrum for qcd from
  ads supergravity duality},  {\em Nucl. Phys.} {\bf B587} (2000) 249--276,
  [\href{http://xxx.lanl.gov/abs/hep-th/0003115}{{\tt hep-th/0003115}}].

\bibitem{Fayyazuddin:1998fb}
A.~Fayyazuddin and M.~Spalinski, {\it Large n superconformal gauge theories and
  supergravity orientifolds},  {\em Nucl. Phys.} {\bf B535} (1998) 219--232,
  [\href{http://xxx.lanl.gov/abs/hep-th/9805096}{{\tt hep-th/9805096}}].

\bibitem{Aharony:1998xz}
O.~Aharony, A.~Fayyazuddin, and J.~M. Maldacena, {\it The large n limit of n =
  2,1 field theories from three- branes in f-theory},  {\em JHEP} {\bf 07}
  (1998) 013, [\href{http://xxx.lanl.gov/abs/hep-th/9806159}{{\tt
  hep-th/9806159}}].

\bibitem{Bertolini:2001qa}
M.~Bertolini, P.~Di~Vecchia, M.~Frau, A.~Lerda, and R.~Marotta, {\it N = 2
  gauge theories on systems of fractional d3/d7 branes},  {\em Nucl. Phys.}
  {\bf B621} (2002) 157--178,
  [\href{http://xxx.lanl.gov/abs/hep-th/0107057}{{\tt hep-th/0107057}}].

\bibitem{Karch:2002sh}
A.~Karch and E.~Katz, {\it Adding flavor to ads/cft},  {\em JHEP} {\bf 06}
  (2002) 043, [\href{http://xxx.lanl.gov/abs/hep-th/0205236}{{\tt
  hep-th/0205236}}].

\bibitem{Sakai:2003wu}
T.~Sakai and J.~Sonnenschein, {\it Probing flavored mesons of confining gauge
  theories by supergravity},  {\em JHEP} {\bf 09} (2003) 047,
  [\href{http://xxx.lanl.gov/abs/hep-th/0305049}{{\tt hep-th/0305049}}].

\bibitem{Nastase:2003dd}
H.~Nastase, {\it On dp-dp+4 systems, qcd dual and phenomenology},
  \href{http://xxx.lanl.gov/abs/hep-th/0305069}{{\tt hep-th/0305069}}.

\bibitem{Ouyang:2003df}
P.~Ouyang, {\it Holomorphic d7-branes and flavored n = 1 gauge theories},  {\em
  Nucl. Phys.} {\bf B699} (2004) 207--225,
  [\href{http://xxx.lanl.gov/abs/hep-th/0311084}{{\tt hep-th/0311084}}].

\bibitem{Nunez:2003cf}
C.~Nunez, A.~Paredes, and A.~V. Ramallo, {\it Flavoring the gravity dual of n =
  1 yang-mills with probes},  {\em JHEP} {\bf 12} (2003) 024,
  [\href{http://xxx.lanl.gov/abs/hep-th/0311201}{{\tt hep-th/0311201}}].

\bibitem{Burrington:2004id}
B.~A. Burrington, J.~T. Liu, L.~A. Pando~Zayas, and D.~Vaman, {\it Holographic
  duals of flavored n = 1 super yang-mills: Beyond the probe approximation},
  {\em JHEP} {\bf 02} (2005) 022,
  [\href{http://xxx.lanl.gov/abs/hep-th/0406207}{{\tt hep-th/0406207}}].

\bibitem{Erdmenger:2004dk}
J.~Erdmenger and I.~Kirsch, {\it Mesons in gauge / gravity dual with large
  number of fundamental fields},  {\em JHEP} {\bf 12} (2004) 025,
  [\href{http://xxx.lanl.gov/abs/hep-th/0408113}{{\tt hep-th/0408113}}].

\bibitem{Hong:2004sa}
S.~Hong, S.~Yoon, and M.~J. Strassler, {\it On the couplings of vector mesons
  in ads/qcd},  \href{http://xxx.lanl.gov/abs/hep-th/0409118}{{\tt
  hep-th/0409118}}.

\bibitem{Kruczenski:2004me}
M.~Kruczenski, L.~A.~P. Zayas, J.~Sonnenschein, and D.~Vaman, {\it Regge
  trajectories for mesons in the holographic dual of large-n(c) qcd},
  \href{http://xxx.lanl.gov/abs/hep-th/0410035}{{\tt hep-th/0410035}}.

\bibitem{Paredes:2004is}
A.~Paredes and P.~Talavera, {\it Multiflavour excited mesons from the fifth
  dimension},  {\em Nucl. Phys.} {\bf B713} (2005) 438--464,
  [\href{http://xxx.lanl.gov/abs/hep-th/0412260}{{\tt hep-th/0412260}}].

\bibitem{Babington:2003vm}
J.~Babington, J.~Erdmenger, N.~J. Evans, Z.~Guralnik, and I.~Kirsch, {\it
  Chiral symmetry breaking and pions in non-supersymmetric gauge / gravity
  duals},  {\em Phys. Rev.} {\bf D69} (2004) 066007,
  [\href{http://xxx.lanl.gov/abs/hep-th/0306018}{{\tt hep-th/0306018}}].

\bibitem{Babington:2003up}
J.~Babington, J.~Erdmenger, N.~J. Evans, Z.~Guralnik, and I.~Kirsch, {\it A
  gravity dual of chiral symmetry breaking},  {\em Fortsch. Phys.} {\bf 52}
  (2004) 578--582, [\href{http://xxx.lanl.gov/abs/hep-th/0312263}{{\tt
  hep-th/0312263}}].

\bibitem{Kruczenski:2003uq}
M.~Kruczenski, D.~Mateos, R.~C. Myers, and D.~J. Winters, {\it Towards a
  holographic dual of large-n(c) qcd},  {\em JHEP} {\bf 05} (2004) 041,
  [\href{http://xxx.lanl.gov/abs/hep-th/0311270}{{\tt hep-th/0311270}}].

\bibitem{Barbon:2004dq}
J.~L.~F. Barbon, C.~Hoyos, D.~Mateos, and R.~C. Myers, {\it The holographic
  life of the eta'},  {\em JHEP} {\bf 10} (2004) 029,
  [\href{http://xxx.lanl.gov/abs/hep-th/0404260}{{\tt hep-th/0404260}}].

\bibitem{Evans:2004ia}
N.~J. Evans and J.~P. Shock, {\it Chiral dynamics from ads space},  {\em Phys.
  Rev.} {\bf D70} (2004) 046002,
  [\href{http://xxx.lanl.gov/abs/hep-th/0403279}{{\tt hep-th/0403279}}].

\bibitem{Ghoroku:2004sp}
K.~Ghoroku and M.~Yahiro, {\it Chiral symmetry breaking driven by dilaton},
  {\em Phys. Lett.} {\bf B604} (2004) 235--241,
  [\href{http://xxx.lanl.gov/abs/hep-th/0408040}{{\tt hep-th/0408040}}].

\bibitem{Bak:2004nt}
D.~Bak and H.-U. Yee, {\it Separation of spontaneous chiral symmetry breaking
  and confinement via ads/cft correspondence},  {\em Phys. Rev.} {\bf D71}
  (2005) 046003, [\href{http://xxx.lanl.gov/abs/hep-th/0412170}{{\tt
  hep-th/0412170}}].

\bibitem{DaRold:2005zs}
L.~Da~Rold and A.~Pomarol, {\it Chiral symmetry breaking from five dimensional
  spaces},  \href{http://xxx.lanl.gov/abs/hep-ph/0501218}{{\tt
  hep-ph/0501218}}.

\bibitem{Evans:2005ti}
N.~Evans, J.~Shock, and T.~Waterson, {\it D7 brane embeddings and chiral
  symmetry breaking},  \href{http://xxx.lanl.gov/abs/hep-th/0502091}{{\tt
  hep-th/0502091}}.

\bibitem{Sakai:2004cn}
T.~Sakai and S.~Sugimoto, {\it Low energy hadron physics in holographic qcd},
  \href{http://xxx.lanl.gov/abs/hep-th/0412141}{{\tt hep-th/0412141}}.

\bibitem{Kruczenski:2003be}
M.~Kruczenski, D.~Mateos, R.~C. Myers, and D.~J. Winters, {\it Meson
  spectroscopy in ads/cft with flavour},  {\em JHEP} {\bf 07} (2003) 049,
  [\href{http://xxx.lanl.gov/abs/hep-th/0304032}{{\tt hep-th/0304032}}].

\bibitem{Karch:2002xe}
A.~Karch, E.~Katz, and N.~Weiner, {\it Hadron masses and screening from ads
  wilson loops},  {\em Phys. Rev. Lett.} {\bf 90} (2003) 091601,
  [\href{http://xxx.lanl.gov/abs/hep-th/0211107}{{\tt hep-th/0211107}}].

\bibitem{Hong:2003jm}
S.~Hong, S.~Yoon, and M.~J. Strassler, {\it Quarkonium from the fifth
  dimension},  {\em JHEP} {\bf 04} (2004) 046,
  [\href{http://xxx.lanl.gov/abs/hep-th/0312071}{{\tt hep-th/0312071}}].

\bibitem{Guralnik:2004ve}
Z.~Guralnik, S.~Kovacs, and B.~Kulik, {\it Holography and the higgs branch of n
  = 2 sym theories},  \href{http://xxx.lanl.gov/abs/hep-th/0405127}{{\tt
  hep-th/0405127}}.

\bibitem{Guralnik:2004wq}
Z.~Guralnik, {\it Strong coupling dynamics of the higgs branch: Rolling a higgs
  by collapsing an instanton},
  \href{http://xxx.lanl.gov/abs/hep-th/0412074}{{\tt hep-th/0412074}}.

\bibitem{Guralnik:2005jg}
Z.~Guralnik, S.~Kovacs, and B.~Kulik, {\it Ads/cft duality and the higgs branch
  of n = 2 sym},  \href{http://xxx.lanl.gov/abs/hep-th/0501154}{{\tt
  hep-th/0501154}}.

\bibitem{Breitenlohner:1982jf}
P.~Breitenlohner and D.~Z. Freedman, {\it Stability in gauged extended
  supergravity},  {\em Ann. Phys.} {\bf 144} (1982) 249.

\bibitem{Skenderis:2002vf}
K.~Skenderis and M.~Taylor, {\it Branes in ads and pp-wave spacetimes},  {\em
  JHEP} {\bf 06} (2002) 025,
  [\href{http://xxx.lanl.gov/abs/hep-th/0204054}{{\tt hep-th/0204054}}].

\bibitem{Douglas:1995bn}
M.~R. Douglas, {\it Branes within branes},
  \href{http://xxx.lanl.gov/abs/hep-th/9512077}{{\tt hep-th/9512077}}.

\bibitem{Witten:1995gx}
E.~Witten, {\it Small instantons in string theory},  {\em Nucl. Phys.} {\bf
  B460} (1996) 541--559, [\href{http://xxx.lanl.gov/abs/hep-th/9511030}{{\tt
  hep-th/9511030}}].

\bibitem{Douglas:1996uz}
M.~R. Douglas, {\it Gauge fields and d-branes},  {\em J. Geom. Phys.} {\bf 28}
  (1998) 255--262, [\href{http://xxx.lanl.gov/abs/hep-th/9604198}{{\tt
  hep-th/9604198}}].

\bibitem{Dorey:2002ik}
N.~Dorey, T.~J. Hollowood, V.~V. Khoze, and M.~P. Mattis, {\it The calculus of
  many instantons},  {\em Phys. Rept.} {\bf 371} (2002) 231--459,
  [\href{http://xxx.lanl.gov/abs/hep-th/0206063}{{\tt hep-th/0206063}}].

\end{thebibliography}\endgroup

\end{document}